\documentstyle[preprint,aps]{revtex}
\begin{document}
\tightenlines
\title{
\begin{flushright}\normalsize HU-EP-01/19\\ \smallskip RCNP-Th 1008\\
\vskip 1cm
\end{flushright}
Rare radiative $\bbox{B}$ decays to orbitally excited
$\bbox{K}$ mesons}
\author{D. Ebert$^{1,2}$,
 R. N. Faustov$^{2,3}$, V. O. Galkin$^{2,3}$ and H. Toki$^{1}$}
\address{$^1$Research Center for Nuclear Physics (RCNP), Osaka University,
Ibaraki, Osaka 567, Japan\\
$^2$Institut f\"ur Physik, Humboldt--Universit\"at
zu Berlin, 10115 Berlin, Germany\\
$^3$Russian Academy of Sciences, Scientific Council for
Cybernetics, Moscow 117333, Russia}
\maketitle
\begin{abstract}
The exclusive rare radiative  $B$ meson decays to orbitally
excited axial-vector mesons  $K_1^*(1270)$,
$K_1(1400)$ and to the tensor meson $K_2^*(1430)$
are investigated in the
framework of the relativistic quark model based on the
quasipotential approach in quantum field theory. These decays are
considered without employing the heavy
quark expansion for the $s$ quark. Instead the
$s$ quark is treated to be light and the expansion in inverse powers
of the large recoil momentum of the final $K^{**}$ meson is used to
simplify calculations. 
It is found that the ratio of the branching fractions of rare radiative
$B$ decays to axial vector $K^*_1(1270)$ and $K_1(1400)$ mesons is
significantly influenced by relativistic effects.  The obtained results for
$B$ decays to the tensor meson $K_2^*(1430)$ agree with recent
experimental data from CLEO.

\medskip
\noindent PACS number(s): 13.20.He, 12.39.Ki
\end{abstract}

\section{Introduction}
\label{sec:int}
Rare radiative decays of $B$ mesons represent an important test of the standard
model of electroweak interactions. These transitions are induced by flavour
changing neutral currents and thus they are sensitive probes of new physics
beyond the standard model.
Such decays are governed by one-loop (penguin) diagrams with
the main contribution from a virtual top quark and a $W$ boson. Therefore,
they provide valuable information about the Cabibbo-Kobayashi-Maskawa
(CKM) matrix elements $V_{ts}$ and $V_{tb}$. The
statistics of rare radiative $B$ decays considerably increased since
the first observation of the $B\to K^*\gamma$ decay in 1993 by  CLEO 
\cite{cleo1}. This allowed a significantly more precise determination
of exclusive and inclusive branching fractions \cite{cleo2}. Recently
the first observation of the rare $B$ decays to the orbitally excited
strange mesons has been reported by CLEO \cite{cleo2}. The branching
fraction for the decay to the tensor $K_2^*(1430)$ meson has been
measured  
$BR(B\to K_2^*(1430)\gamma)=(1.66^{+0.59}_{-0.53}\pm 0.13)\times 10^{-5},$
as well as the ratio of exclusive branching fractions
$r\equiv
{BR(B\to K_2^*(1430)\gamma )}/{BR(B\to K^*(892)\gamma)}
=0.39^{+0.15}_{-0.13}.$
The data for the other decay channels will be available soon.
This significant experimental progress provides a challenge to the theory.
Many theoretical approaches have been employed to predict the
exclusive $B\to K^*(892)\gamma$ decay rate (for a review see
\cite{lss} and references therein). Considerably less attention has been payed
to rare radiative $B$ decays to excited strange mesons
\cite{a,aom,as,vo}. Most of these theoretical approaches
\cite{aom,vo}  rely on the heavy quark limit both for the initial
$b$ and final $s$ quarks and the nonrelativistic quark model.
However, the two predictions \cite{aom,vo} for the ratio $r$ 
differ by an order of magnitude, due to a
different treatment of the long distance effects and, as a
result, a different determination of corresponding Isgur-Wise
functions. 
Only the prediction of Ref.~\cite{vo} is consistent
with the available data. Nevertheless, it is
necessary to point out that the $s$ quark in the final $K^*$
meson is not heavy enough, compared to the $\bar \Lambda$
parameter, which determines the scale of $1/m_Q$ corrections in
heavy quark effective theory \cite{n}. Thus the $1/m_s$ expansion
is not appropriate. Notwithstanding, the ideas of heavy quark
expansion can be applied to the exclusive $B\to
K^*(K^{**})\gamma$  decays. From the kinematical analysis it
follows that the final $K^*(K^{**})$ meson bears a large
relativistic recoil momentum $\vert {\bf\Delta} \vert$ of order
of $m_b/2$ and an energy of the same order. So it is possible to
expand the matrix element of the effective Hamiltonian both in
inverse powers of the $b$ quark mass for the initial state and in
inverse powers of the recoil momentum $\vert{\bf\Delta} \vert$
for the final state \cite{gf,cbf}. Such an expansion has been realized by us
for the $B\to K^*(892)\gamma$ decay in the framework of the
relativistic quark model \cite{gf}.
In Refs.~\cite{cbf} it was shown that in the leading order of this
expansion a specific symmetry emerges which imposes several relations
between the form factors of semileptonic and rare radiative $B$ decays.
{It is important to
note that rare radiative decays of $B$ mesons require a
completely relativistic treatment, because the recoil momentum of
the final meson is large compared to the $s$ quark mass.} The
calculated branching fraction for this decay \cite{gf} was found in good
agreement with experimental data. In Ref.~\cite{efgt} we considered
the exclusive rare $B$ decay to the orbitally excited tensor
meson $K_2^*(1430)$. Here we extend this analysis to the axial-vector
mesons $K^*_1(1270)$ and $K_1(1400)$.

Our relativistic quark model is  based on the quasipotential approach in
quantum field
theory with a specific choice  of the quark-antiquark interaction
potential. It provides
a consistent scheme for the  calculation of all relativistic corrections
at a given $v^2/c^2$
order and allows for the heavy  quark $1/m_Q$ expansion. In preceding papers
we applied this model to the calculation of the mass spectra of
orbitally and radially excited states of heavy-light mesons \cite{egf},
as well as to the description of weak decays of $B$ mesons to ground state
heavy and light mesons \cite{fgm,efg}. The heavy quark expansion for the
heavy-to-heavy semileptonic transitions \cite{fg,radexc} was found to be
in agreement with model-independent predictions of the heavy quark effective
theory (HQET).

The  paper is organized as follows. In Sec.~II we define the form
factors, which govern the exclusive rare radiative $B$ decays to
orbitally excited $K$ mesons.
The relativistic quark model is described in Sec.~III, and in Sec.~IV
it is applied for the calculation of the rare radiative decay form
factors.  Our numerical results for the form factors and decay rates
as well as  comparison of these results  with other theoretical
predictions and experimental data are presented in Sec.~V. There we
also discuss the relations between the form factors of rare radiative
and semileptonic $B$ decays  in the formal heavy quark
limit.  Sec.~VI contains our conclusions.

\section{Rare radiative $\bbox{B}$ decays }
\label{sec:rr}
In the standard model, $B$ decays are described by the effective
Hamiltonian, after integrating out the top quark and $W$ boson
and using the Wilson expansion \cite{eh}. For the case of $b \to s$
transition:
\begin{equation}\label{heff}
H_{eff}(b\to s)=-\frac{4G_F}{\sqrt2}
V^*_{ts}V_{tb}\sum ^8_{j=1} C_j(\mu)O_j(\mu),
\end{equation}
where $V_{ij}$ are the corresponding  CKM
matrix elements, $ \lbrace
O_j\rbrace $ are a complete set of renormalized dimension six
operators involving light fields, which govern $b\to s$ transitions.
They include six four-quark operators $O_j\quad (j=1,\ldots ,6)$,
which determine the non-leptonic $B$ decay rates, the electromagnetic
dipole operator
\begin{equation}\label{o7}
O_7=\frac{e}{16\pi ^2}\bar s\sigma ^{\mu \nu
}(m_bP_R+m_sP_L)bF_{\mu \nu }, \qquad P_{R,L}=(1 \pm \gamma _5)/2,
\end{equation}
and the chromomagnetic dipole operator $O_8$.
$O_7$ and $O_8$ are responsible for the rare $B$ decays 
$b\to s\gamma$ and $b \to sg$,
respectively \cite{eh}. The Wilson coefficients $C_j(\mu )$ are evaluated
perturbatively at the $W$ scale and then they are evolved down to the
renormalization scale $\mu \sim m_b$ by the renormalization group
equations. There are also power-suppressed terms $\sim 1/m_c^2$ \cite{krsw,b}.

The dominant contribution to $B\to K^{**}\gamma$ decay rates come from
the electromagnetic dipole operator $O_7$.
The matrix elements of this operator between the initial $B$ meson state and
the final state of the orbitally
excited $K^{**}$ meson have the following covariant decomposition
\begin{eqnarray}\label{ff1}
\langle K_1(p',\epsilon)|\bar s i k_\nu \sigma_{\mu \nu}
b|B(p)\rangle & = &
t_{+}(k^2)\left((\epsilon^*\cdot k)(p+p')_\mu -
\epsilon_{\mu}^*(p^2-p'^2)  \right) \cr
&& +t_{-}(k^2)\left((\epsilon^*\cdot k) k_\mu
-\epsilon_{\mu}^* k^2\right) \cr
& &+t_0(k^2)(\epsilon^*\cdot k)((p^2-p'^2)k_\mu -(p+p')_\mu k^2)/(M_B 
M_{K_1}),\cr
\langle K_1(p',\epsilon)|\bar s i k_\nu \sigma_{\mu \nu}\gamma_5 b|B(p)\rangle
&=&i t_{+}(k^2)
\epsilon_{\mu \nu \lambda \sigma } \epsilon^{*\nu}
k^\lambda (p+p')^\sigma , \\
\label{ff1*}
\langle K_1^*(p',\epsilon)|\bar s i k_\nu \sigma_{\mu \nu}
b|B(p)\rangle & = &
s_{+}(k^2)\left(  (\epsilon^*\cdot k)(p+p')_\mu -
\epsilon_{\mu}^*(p^2-p'^2)\right)\cr
& & +s_{-}(k^2)\left((\epsilon^*\cdot k) k_\mu
-\epsilon_{\mu}^* k^2\right) \cr
& &+s_0(k^2)(\epsilon^*\cdot k)((p^2-p'^2)k_\mu -(p+p')_\mu k^2)/(M_B
M_{K_1^*}),\cr
\langle K_1^*(p',\epsilon)|\bar s i k_\nu \sigma_{\mu \nu}\gamma_5 b|B(p)\rangle
&=&i s_{+}(k^2)
\epsilon_{\mu \nu \lambda \sigma } \epsilon^{*\nu}
k^\lambda (p+p')^\sigma , \\
\label{ff2}
\langle K_2^*(p',\epsilon)|\bar s i k_\nu \sigma_{\mu \nu}b|B(p)\rangle
&=&i g_{+}(k^2)
\epsilon_{\mu \nu \lambda \sigma } \epsilon^{*\nu \beta}\frac{p_\beta}{M_B}
k^\lambda (p+p')^\sigma , \cr
\langle K_2^*(p',\epsilon)|\bar s i k_\nu \sigma_{\mu \nu}
\gamma_5b|B(p)\rangle & = &
g_{+}(k^2)\left( \epsilon_{\beta \gamma}^*\frac{p^\beta
p^\gamma }{M_B}(p+p')_\mu -
\epsilon_{\mu \beta}^*\frac{p^\beta }{M_B}(p^2-p'^2)
 \right) \cr
& & +g_{-}(k^2)\left(\epsilon_{\beta \gamma}^*\frac{p^\beta
p^\gamma }{M_B}k_\mu
-\epsilon_{\mu \beta}^*\frac{p^\beta }{M_B}k^2\right) \cr
& &+g_0(k^2)((p^2-p'^2)k_\mu -(p+p')_\mu k^2)\epsilon_{\beta\gamma }^*
\frac{p^\beta p^\gamma  }{M_B^2 M_{K_2^*}},
\end{eqnarray}
where $\epsilon_\mu (\epsilon_{\mu\nu})$ is a polarization vector
(tensor) of the final axial-vector (tensor) meson and $k= p -p'$ is
the four momentum of the emitted photon. The exclusive decay rates for
the emission of a real photon ($k^2=0$) are determined by form factors
$t_{+}(0)$, $s_{+}(0)$ and $g_{+}(0)$. They  are given by
\begin{eqnarray}
\label{drate1}
\Gamma(B\to K_1\gamma)&=&
\frac{\alpha }{32\pi^4} G_F^2m_b^5|V_{tb}V_{ts}|^2
|C_7(m_b)|^2 t_{+}^2(0)
\left(1-\frac{M_{K_1}^2}{M_B^2}
 \right)^3\left( 1+\frac{M_{K_1}^2}{M_B^2}\right),\\
\label{drate11}
\Gamma(B\to K_1^*\gamma)&=&
\frac{\alpha }{32\pi^4} G_F^2m_b^5|V_{tb}V_{ts}|^2
|C_7(m_b)|^2 s_{+}^2(0)
\left(1-\frac{M_{K_1^*}^2}{M_B^2}
 \right)^3\left( 1+\frac{M_{K_1^*}^2}{M_B^2}\right),\\
\label{drate2}
\Gamma(B\to K_2^*\gamma)&=&
\frac{\alpha }{256\pi^4} G_F^2m_b^5|V_{tb}V_{ts}|^2
|C_7(m_b)|^2 g_{+}^2(0) \frac{M_B^2}{M_{K_2^*}^2}
\left(1-\frac{M_{K_2^*}^2}{M_B^2}
 \right)^5\left( 1+\frac{M_{K_2^*}^2}{M_B^2}\right),
\end{eqnarray}
where $C_7(m_b)$ is the Wilson coefficient in front of the operator $O_7$.
It is convenient to consider the ratio of exclusive to inclusive
branching fractions, for which we have
\begin{eqnarray}
\label{rk1}
R_{K_1}&\equiv&
\frac{BR(B\to K_1(1400)\gamma)}{BR(B\to X_s\gamma)}=
 t_{+}^2(0)
\frac{\left(1-{M_{K_1}^2}/{M_B^2}
 \right)^3\left( 1+{M_{K_1}^2}/{M_B^2}\right)}{\left(1-{m_s^2}/{m_b^2}
 \right)^3\left( 1+{m_s^2}/{m_b^2}\right)},\\
\label{rk2}
R_{K_1^*}&\equiv&
\frac{BR(B\to K_1^*(1270)\gamma)}{BR(B\to X_s\gamma)}=
s_{+}^2(0)
\frac{\left(1-{M_{K_1^*}^2}/{M_B^2}
 \right)^3\left( 1+{M_{K_1^*}^2}/{M_B^2}\right)}{\left(1-{m_s^2}/{m_b^2}
 \right)^3\left( 1+{m_s^2}/{m_b^2}\right)},\\
R_{K_2^*}&\equiv&
\frac{BR(B\to K_2^*(1430)\gamma)}{BR(B\to X_s\gamma)}=
\frac18 g_{+}^2(0)\frac{M_B^2}{M_{K_2^*}^2}
\frac{\left(1-{M_{K_2^*}^2}/{M_B^2}
 \right)^5\left( 1+{M_{K_2^*}^2}/{M_B^2}\right)}{\left(1-{m_s^2}/{m_b^2}
 \right)^3\left( 1+{m_s^2}/{m_b^2}\right)}.
\label{rk11}
\end{eqnarray}
The recent experimental value for the inclusive decay branching
fraction \cite{t}
$$BR(B\to X_s\gamma)=(3.15\pm 0.35\pm 0.32\pm 0.26)\times 10^{-4}$$
is in a good agreement with theoretical calculations (see e.g. \cite{lss}).

\section{Relativistic quark model}
\label{sec:rqm}

Now we use the relativistic quark model for the calculation of the form factors
$t_{+}(0)$, $s_{+}(0)$ and
$g_{+}(0)$. In our model a meson is described by the wave
function of the bound quark-antiquark state, which satisfies the
quasipotential equation \cite{3} of the Schr\"odinger type~\cite{4} in
the center-of-mass frame:
\begin{equation}
\label{quas}
{\left(\frac{b^2(M)}{2\mu_{R}}-\frac{{\bf
p}^2}{2\mu_{R}}\right)\Psi_{M}({\bf p})} =\int\frac{d^3 q}{(2\pi)^3}
 V({\bf p,q};M)\Psi_{M}({\bf q}),
\end{equation}
where the relativistic reduced mass
is
\begin{equation}
\mu_{R}=\frac{M^4-(m^2_q-m^2_Q)^2}{4M^3},
\end{equation}
and 
$b^2(M)$  denotes
the on-mass-shell relative momentum squared
\begin{equation}
{b^2(M) }
=\frac{[M^2-(m_q+m_Q)^2][M^2-(m_q-m_Q)^2]}{4M^2}.
\end{equation}

The kernel
$V({\bf p,q};M)$ in Eq.~(\ref{quas}) is the quasipotential operator of
the quark-antiquark interaction. It is constructed with the help of the
off-mass-shell scattering amplitude, projected onto the positive
energy states. An important role in this construction is played
by the Lorentz-structure of the confining quark-antiquark interaction
in the meson.  In
constructing the quasipotential of the quark-antiquark interaction
we have assumed that the effective
interaction is the sum of the usual one-gluon exchange term and the mixture
of vector and scalar linear confining potentials.
The quasipotential is then defined by
\cite{mass}
\begin{equation}
\label{qpot}
V({\bf p,q};M)=\bar{u}_q(p)\bar{u}_Q(-p){\cal V}({\bf p}, {\bf
q};M)u_q(q)u_Q(-q),
\end{equation}
with
$${\cal V}({\bf p},{\bf q};M)=\frac{4}{3}\alpha_sD_{ \mu\nu}({\bf
k})\gamma_q^{\mu}\gamma_Q^{\nu}
+V^V_{\rm conf}({\bf k})\Gamma_q^{\mu}
\Gamma_{Q;\mu}+V^S_{\rm conf}({\bf k}),$$
where $\alpha_s$ is the QCD coupling constant, $D_{\mu\nu}$ is the
gluon propagator in the Coulomb gauge
and ${\bf k=p-q}$; $\gamma_{\mu}$ and $u(p)$ are
the Dirac matrices and spinors
\begin{equation}
\label{spinor}
u^\lambda({p})=\sqrt{\frac{\epsilon(p)+m}{2\epsilon(p)}}
\left(\begin{array}{c}
1\\ \displaystyle\frac{\mathstrut\bbox{\sigma}{\bf p}}{\mathstrut\epsilon(p)+m}
\end{array}\right)
\chi^\lambda
\end{equation}
with $\epsilon(p)=\sqrt{{\bf p}^2+m^2}$.
The effective long-range vector vertex is
given by
\begin{equation}
\Gamma_{\mu}({\bf k})=\gamma_{\mu}+
\frac{i\kappa}{2m}\sigma_{\mu\nu}k^{\nu},
\end{equation}
where $\kappa$ is the Pauli interaction constant characterizing the
nonperturbative anomalous chromomagnetic moment of quarks. Vector and
scalar confining potentials in the nonrelativistic limit reduce to
\begin{equation}\label{vconf}
V^V_{\rm conf}(r)=(1-\varepsilon)(Ar+B),\qquad
V^S_{\rm conf}(r) =\varepsilon (Ar+B),
\end{equation}
reproducing
\begin{equation}
V_{\rm conf}(r)=V^S_{\rm conf}(r)+
V^V_{\rm conf}(r)=Ar+B,
\end{equation}
where $\varepsilon$ is the mixing coefficient.

The quasipotential for the heavy quarkonia,
expanded in $v^2/c^2$, can be found in Refs.~\cite{mass,pot} and for
heavy-light mesons in \cite{egf}.
All the parameters of
our model, such as quark masses, parameters of the linear confining potential,
mixing coefficient $\varepsilon$ and anomalous
chromomagnetic quark moment $\kappa$, were fixed from the analysis of
heavy quarkonia masses \cite{mass} and radiative decays \cite{gfm}.
The quark masses
$m_b=4.88$ GeV, $m_c=1.55$ GeV, $m_s=0.50$ GeV, $m_{u,d}=0.33$ GeV and
the parameters of the linear potential $A=0.18$ GeV$^2$ and $B=-0.30$ GeV
have the usual quark model values.
In Ref.~\cite{fg} we have considered the expansion of  the matrix
elements of weak heavy quark currents between pseudoscalar and vector
meson ground states up to the second order in inverse powers of
the heavy quark
masses. It has been found that the general structure of the leading,
first,
and second order $1/m_Q$ corrections in our relativistic model is in accord
with the predictions of HQET. The heavy quark symmetry and QCD impose rigid
constraints on the parameters of the long-range potential in our model.
The analysis
of the first order corrections \cite{fg} fixes the value of the
Pauli interaction
constant $\kappa=-1$. The same value of $\kappa$  was found previously
from  the fine splitting of heavy quarkonia ${}^3P_J$- states
\cite{mass}.
The value of the parameter characterizing the  
 mixing of
vector and scalar confining potentials,
 $\varepsilon=-1$,
was found from the analysis of the second order corrections \cite{fg}.
This value is very close to the one determined from considering radiative
decays of heavy quarkonia \cite{gfm}.

\section{Rare radiative $\bbox{B\to K^{**}\gamma}$ decay form factors}
\label{sec:dff}

In the quasipotential approach,  the matrix element of the weak
current $J_\mu=\bar s \frac{i}{2} k^\nu
\sigma_{\mu\nu}(1+\gamma^5)b$ between the states of a $B$ meson
and an orbitally excited $K^{**}$ meson has the form \cite{f}
\begin{equation}\label{mxet}
\langle K^{**} \vert J_\mu (0) \vert B\rangle
=\int \frac{d^3p\, d^3q}{(2\pi )^6} \bar \Psi_{K^{**}}({\bf
p})\Gamma _\mu ({\bf p},{\bf q})\Psi_B({\bf q}),\end{equation}
where $\Gamma _\mu ({\bf p},{\bf
q})$ is the two-particle vertex function and  $\Psi_{B,K^{**}}$ are the
meson wave functions projected onto the positive energy states of
quarks and boosted to the moving reference frame.
 The contributions to $\Gamma$ come from Figs.~1 and 2.
The contribution $\Gamma^{(2)}$ is the consequence of the
projection onto the positive-energy states. Note that the form
of the relativistic corrections resulting from the vertex function
$\Gamma^{(2)}$ explicitly depends on the Lorentz structure of the
$q\bar q$-interaction.  The vertex functions look like
\begin{equation}\label{gam1}
\Gamma_\mu ^{(1)}({\bf p},{\bf q})=\bar
u_s(p_1)\frac{i}{2}\sigma_{\mu \nu} k^\nu
(1+\gamma^5)u_b(q_1)(2\pi)^3\delta({\bf p}_2-{\bf q}_2),\end{equation}
and
\begin{eqnarray}\label{gam2}
\Gamma_\mu^{(2)}({\bf p},{\bf q})&=&\bar u_s(p_1)\bar
u_q(p_2)\frac{1}{ 2} \biggl\{ i\sigma_{1\mu
\nu}k_\nu(1+\gamma_1^5)\frac{\Lambda_b^{(-)}({k}_1)}{ \epsilon
_b(k_1)+\epsilon_b(p_1)}\gamma_1^0{\cal V}({\bf p}_2-{\bf q}_2)\nonumber\\
& & +{\cal V}({\bf p}_2-{\bf q}_2)\frac{\Lambda_s^{(-)}(k_1')}{
\epsilon_s(k_1')+ \epsilon_s(q_1)}\gamma_1^0i\sigma_{1\mu
\nu}k_\nu(1+\gamma_1^5)\biggr\}u_b(q_1) u_q(q_2), \end{eqnarray}
where ${\bf k}_1={\bf p}_1-{\bf\Delta};\quad {\bf k}_1'={\bf
q}_1+{\bf\Delta};\quad {\bf\Delta}={\bf p}_{K^{**}}-{\bf p}_B$;
$$\Lambda^{(-)}(p)={\epsilon(p)-\bigl( m\gamma ^0+\gamma^0({\bf
\bbox{\gamma} p})\bigr) \over 2\epsilon (p)}.$$

The wave functions of $P$-wave $K^{**}$ mesons at rest can be
parametrized either through the wave functions of the states $^1P_1$,
$^3P_{0,1,2}$ used in quark models for quarkonia ($LS$-coupling scheme)
or through the wave functions $K(1/2)$ and $K(3/2)$ used in HQET
($js$-coupling scheme). The structure of the wave
functions for the states with $0^+$ and $2^+$ quantum numbers is the
same in both parametrizations, while two real states with $1^+$ quantum
numbers are different mixtures of states in these
parametrization. Experiment shows that $K_1(1400)$ and $K^*_1(1270)$
mesons are nearly equal mixtures of $^1P_1$ and $^3P_1$ quark model
states \cite{pdg}. As a result the HQET parametrization turns out to be
more appropriate since the real $K_1(1400)$ and $K^*_1(1270)$ mesons
almost coincide with the corresponding states in $js$-coupling
scheme. The wave functions at rest in HQET parametrization are given by
\begin{equation}\label{psi}
\Psi_{K^{**}}({\bf p})\equiv
\Psi^{JM}_{K(j)}({\bf p})={\cal Y}^{JM}_j\psi_{K(j)}({\bf p}),
\end{equation}
where $J$ and $M$ are the meson total angular momentum and its projection,
while $j$ is the $u,d$-quark angular momentum;
$\psi_{K(j)}({\bf p})$ is the radial part of the wave function.
The spin-angular momentum part ${\cal Y}^{JM}_j$ has the following
form
\begin{eqnarray}\label{angl}
{\cal Y}^{JM}_j&=&\sum_{\sigma_Q\sigma_q}\left\langle j\, M-\sigma_Q,\
\frac12\, \sigma_Q |J\, M\right\rangle\left\langle 1\, M-\sigma_Q-\sigma_q,\
\frac12\, \sigma_q |j\, M-\sigma_Q\right\rangle \cr \cr
 & &\times Y_{1}^{M-\sigma_Q-\sigma_q}
\chi_Q(\sigma_Q)\chi_q(\sigma_q).
\end{eqnarray}
Here $\langle j_1\, m_1,\  j_2\, m_2|J\, M\rangle$ are the Clebsch-Gordan
coefficients, $Y_l^m$ are the spherical harmonics, and $\chi(\sigma)$ (where
$\sigma=\pm 1/2$) are spin wave functions :
$$ \chi\left(1/2\right)={1\choose 0}, \qquad
\chi\left(-1/2\right)={0\choose 1}. $$
 Then
\begin{eqnarray}
  \label{eq:mix}
  \Psi^M_{K_1(1400)}&=&\Psi^{1M}_{K(3/2)}\cos\phi
  +\Psi^{1M}_{K(1/2)}\sin\phi,\cr
 \Psi^M_{K^*_1(1270)}&=&\Psi^{1M}_{K(1/2)}\cos\phi
  -\Psi^{1M}_{K(3/2)}\sin\phi,
\end{eqnarray}
where $\phi$ is a small mixing angle. We have calculated the wave
functions of orbitally excited $K^{**}$ mesons in our model by the
numerical solution of Eq.~(\ref{quas}) with the quasipotential
(\ref{qpot}) expanded in inverse powers of quark energies. Such an
expansion is more adequate for $K$ mesons than the usual
nonrelativistic expansion since the velocities of light $u,d,s$ quarks
are highly relativistic. The calculated spin-averaged
$P$-wave $K$ meson masses as well as spin-orbit splittings are
consistent with experimental values. The obtained value of the mixing
angle in (\ref{eq:mix})  also agrees  with experiment and is
approximately equal to $\phi\approx 4\raisebox{1ex}{\scriptsize o}$.
In the following we will use the functions (\ref{psi}) for decay form
factor calculations assuming that the physical form factors for $B\to
K_1^{(*)}\gamma$ decays are related to the calculated ones
(denoted by a tilde) by
\begin{eqnarray}
  \label{eq:fff}
  t_{+}&=&\tilde t_{+}\cos\phi
  +\tilde s_{+}\sin\phi,\cr
s_{+}&=&\tilde s_{+}\cos\phi
  -\tilde t_{+}\sin\phi.
\end{eqnarray}

It is important to note that the wave functions entering the weak current
matrix element (\ref{mxet}) cannot  be both in the rest frame.
In the $B$ meson rest frame, the $K^{**}$ meson is moving with the recoil
momentum ${\bf \Delta}$. The wave function
of the moving $K^{**}$ meson $\Psi_{K^{**}\,{\bf\Delta}}$ is connected
with the $K^{**}$ wave function in the rest frame
$\Psi_{K^{**}\,{\bf 0}}\equiv \Psi_{K^{**}}$ by the transformation \cite{f}
\begin{equation}
\label{wig}
\Psi_{K^{**}\,{\bf\Delta}}({\bf p})
=D_s^{1/2}(R_{L_{\bf\Delta}}^W)D_q^{1/2}(R_{L_{\bf\Delta}}^W)
\Psi_{K^{**}\,{\bf 0}}({\bf p}),
\end{equation}
where $R^W$ is the Wigner rotation, $L_{\bf\Delta}$ is the Lorentz boost
from the meson rest frame to a moving one, and
the rotation matrix $D^{1/2}(R)$ in spinor representation is given by
\begin{equation}\label{d12}
{1 \ \ \,0\choose 0 \ \ \,1}D^{1/2}_{s,q}(R^W_{L_{\bf\Delta}})=
S^{-1}({\bf p}_{s,q})S({\bf\Delta})S({\bf p}),
\end{equation}
where
$$
S({\bf p})=\sqrt{\frac{\epsilon(p)+m}{2m}}\left(1+\frac{\bbox{\alpha}{\bf p}}
{\epsilon(p)+m}\right)
$$
is the usual Lorentz transformation matrix of the four-spinor.

We substitute the vertex functions $\Gamma^{(1)}$  and $\Gamma^{(2)}$
given by Eqs.~(\ref{gam1}) and (\ref{gam2})
in the decay matrix element (\ref{mxet})  and take into account the wave
function transformation (\ref{wig}).
The resulting structure of this matrix element is
rather complicated, because it is
necessary  to integrate both over  $d^3 p$
and $d^3 q$. The $\delta$ function in expression  (\ref{gam1}) permits
to perform one of these integrations
and thus this contribution  can be easily
calculated. The calculation  of the vertex function
$\Gamma^{(2)}$ contribution is  more difficult. Here, instead
of a $\delta$ function, we have a complicated structure, containing the
$q\bar q$ interaction operator ${\cal V}$.
However, we can expand this contribution in the inverse
powers of the heavy $b$ quark mass and large
recoil momentum $|{\bf \Delta}|\sim
m_b/2$ of the final $K^{**}$ meson.   Such an
expansion is carried out up to the second order.~\footnote{This means that
in expressions for $\tilde t,\tilde s,g_{+}^{(2)V}(0)$ and $\tilde
t,\tilde s,g_{+}^{(2)S}(0)$ we neglect terms
proportional to the third order product of small binding energies and ratios
${\bf p}^2/\epsilon_s^3(\Delta)$, ${\bf p}^2/\epsilon_b^3(\Delta)$
as well as higher order terms.} Then we use the  quasipotential equation in
order to perform one of the integrations in the current matrix element.
As a result we get for the form factors the following expressions
with $\kappa=-1$
\begin{eqnarray}
\label{tpl}
\tilde t_{+}(0)&=&\tilde t_{+}^{(1)}(0)+
(1-\varepsilon)\tilde t_{+}^{(2)V}(0)+\varepsilon\tilde t_{+}^{(2)S}(0),\\
\label{t1}
\tilde t_{+}^{(1)}(0) &=&\frac{1}{3\sqrt{2}} \sqrt{\frac{E_{K(3/2)}}{M_B}}
\frac{|{\bf \Delta}|}{E_{K(3/2)}+
M_{K(3/2)}} \int \frac{d^3p}{(2\pi )^3} \bar\psi_{K(3/2)}
\Bigl({\bf p}+
\frac{2\epsilon_q }{E_{K(3/2)}+M_{K(3/2)}}{\bf \Delta } \Bigr)\cr \cr
&&\times\sqrt{\frac{\epsilon_s(p+\Delta )+m_s}{2\epsilon_s(p+\Delta )}}
\sqrt{\frac{\epsilon_b(p )+m_b}{2\epsilon_b(p )}}
\Biggl\{-3(E_{K(3/2)}+M_{K(3/2)})
\frac{({\bf p}\cdot {\bf\Delta})}{p{\bf \Delta}^2}\cr \cr
&&\times\left[1+\frac{M_B-E_{K(3/2)}}
{\epsilon_s(p+\Delta )+m_s } \right]
+\left[\frac{p}{\epsilon_q(p)+m_q }
-\frac{p}{\epsilon_s(p+\Delta )+m_s} \right] \cr \cr
&&\times\left[1+\frac{M_B-E_{K(3/2)}}
{\epsilon_s(p+\Delta )+m_s }
+\frac{{\bf p}^2}{[\epsilon_s(p+\Delta )+m_s]
[\epsilon_b(p )+m_b]} \right]\cr\cr
&&-2\frac{M_B+M_{K(3/2)}}{M_B-M_{K(3/2)}}
\frac{p}{\epsilon_s(p+\Delta )+m_s}\Biggr\} \psi_B({\bf p}),\\  \cr
\label{t2v}
\tilde t_{+}^{(2)V}(0) &=&\frac{1}{3\sqrt{2}} \sqrt{\frac{E_{K(3/2)}}{M_B}}
\frac{|{\bf \Delta}|}{E_{K(3/2)}+
M_{K(3/2)}} \int \frac{d^3p}{(2\pi )^3} \bar\psi_{K(3/2)}
\Bigl({\bf p}+\frac{2\epsilon_q }{E_{K(3/2)}+
M_{K(3/2)}}{\bf \Delta } \Bigr)\cr \cr
&&\times\sqrt{\frac{\epsilon_s(\Delta )+m_s}{2\epsilon_s(\Delta )}}
\Biggl\{\left[3(E_{K(3/2)}+M_{K(3/2)})
\frac{({\bf p}\cdot {\bf\Delta})}{p{\bf \Delta}^2}
-\frac{p}{\epsilon_q(p)+m_q }\right]\cr \cr
&&\times\frac{\epsilon_s(\Delta )-m_s}
{2\epsilon_s(\Delta )[\epsilon_s(\Delta )+m_s]}
\biggl(M_B-\epsilon_b(p)-\epsilon_q(p)
+\frac{M_B-E_{K(3/2)}}{\epsilon_s(\Delta )+m_s}
\biggl[M_{K(3/2)}\cr \cr
&&-
\epsilon_s\Bigl({\bf p}+\frac{
2\epsilon_q }{E_{K(3/2)}+M_{K(3/2)}}{\bf \Delta }
\Bigr)- \epsilon_s\Bigl({\bf p}+
\frac{2\epsilon_q }{E_{K(3/2)}+M_{K(3/2)}}{\bf
\Delta } \Bigr) \biggr]\biggr)\cr\cr
&&+\frac{p}{\epsilon_q(p)+m_q }\Biggl[
\frac{M_B-\epsilon_b(p)-\epsilon_q(p)}{\epsilon_b(\Delta )+m_b}
+3\frac{\epsilon_s(\Delta )-m_s}
{2\epsilon_s(\Delta )[\epsilon_s(\Delta )+m_s]}\bigglb(M_{K(3/2)}\cr\cr
&&-
\epsilon_s\Bigl({\bf p}+\frac{
2\epsilon_q }{E_{K(3/2)}+M_{K(3/2)}}{\bf \Delta }
\Bigr)- \epsilon_s\Bigl({\bf p}+
\frac{2\epsilon_q }{E_{K(3/2)}+M_{K(3/2)}}{\bf
\Delta } \Bigr)\biggrb) \Biggr]
\Biggr\} \psi_B({\bf p}),\\  \cr
\label{t2s}
\tilde t_{+}^{(2)S}(0) &=&\frac{1}{3\sqrt{2}} \sqrt{\frac{E_{K(3/2)}}{M_B}}
\frac{|{\bf \Delta}|}{E_{K(3/2)}+M_{K(3/2)}}
\int \frac{d^3p}{(2\pi )^3} \bar\psi_{K(3/2)}
\Bigl({\bf p}+
\frac{2\epsilon_q }{E_{K(3/2)}+M_{K(3/2)}}{\bf \Delta } \Bigr)\cr \cr
&&\times\sqrt{\frac{\epsilon_s(\Delta )+m_s}{2\epsilon_s(\Delta )}}
\Biggl\{\Biggl[-3(E_{K(3/2)}+M_{K(3/2)})
\frac{({\bf p}\cdot {\bf\Delta})}{p{\bf \Delta}^2}+
\frac{p}{\epsilon_q(p)+m_q }\Biggr]\cr \cr
&&\times\frac{\epsilon_s(\Delta )-m_s}
{2\epsilon_s(\Delta )[\epsilon_s(\Delta )+m_s]}\Biggl[M_B-\epsilon_b(p)-\epsilon_q(p)
-\frac{M_B-E_{K(3/2)}}{\epsilon_s(\Delta )+m_s}
 \bigglb(M_{K(3/2)}\cr\cr
&&-
\epsilon_s\Bigl({\bf p}+\frac{
2\epsilon_q }{E_{K(3/2)}+M_{K(3/2)}}{\bf \Delta }
\Bigr)-  \epsilon_s\Bigl({\bf p}+
\frac{2\epsilon_q }{E_{K(3/2)}+M_{K(3/2)}}{\bf
\Delta } \Bigr) \biggrb)
\Biggr]\Biggr\} \psi_B({\bf p}),\\
\label{spl}
\tilde s_{+}(0)&=&\tilde s_{+}^{(1)}(0)+
(1-\varepsilon)\tilde s_{+}^{(2)V}(0)+\varepsilon\tilde s_{+}^{(2)S}(0),\\
\label{s1}
\tilde s_{+}^{(1)}(0) &=&\frac{1}{3} \sqrt{\frac{E_{K(1/2)}}{M_B}}
\frac{|{\bf \Delta}|}{E_{K(1/2)}+
M_{K(1/2)}} \int \frac{d^3p}{(2\pi )^3} \bar\psi_{K(1/2)}
\Bigl({\bf p}+
\frac{2\epsilon_q }{E_{K(1/2)}+M_{K(1/2)}}{\bf \Delta } \Bigr)\cr \cr
&&\times\sqrt{\frac{\epsilon_s(p+\Delta )+m_s}{2\epsilon_s(p+\Delta )}}
\sqrt{\frac{\epsilon_b(p )+m_b}{2\epsilon_b(p )}}
\Biggl\{-3(E_{K(1/2)}+M_{K(1/2)})
\frac{({\bf p}\cdot {\bf\Delta})}{p{\bf \Delta}^2}\cr \cr
&&\times\left[1+\frac{M_B-E_{K(1/2)}}
{\epsilon_s(p+\Delta )+m_s } \right]
-\left[\frac{2p}{\epsilon_q(p)+m_q }
-\frac{2p}{\epsilon_s(p+\Delta )+m_s} \right] \cr \cr
&&\times\left[1+\frac{M_B-E_{K(1/2)}}
{\epsilon_s(p+\Delta )+m_s }
+\frac{{\bf p}^2}{[\epsilon_s(p+\Delta )+m_s]
[\epsilon_b(p )+m_b]} \right]
-\frac{M_B+M_{K(1/2)}}{M_B-M_{K(1/2)}}\cr\cr
&&\times \left[\frac{p}{\epsilon_b(p )+m_b}
\left(1+\frac{M_B-E_{K(1/2)}}
{\epsilon_s(p+\Delta )+m_s } \right) +
\frac{3p}{\epsilon_s(p+\Delta )+m_s}\right]\Biggr\} \psi_B({\bf p}),\\  \cr
\label{s2v}
\tilde s_{+}^{(2)V}(0) &=&\frac{1}{3} \sqrt{\frac{E_{K(1/2)}}{M_B}}
\frac{|{\bf \Delta}|}{E_{K(1/2)}+
M_{K(1/2)}} \int \frac{d^3p}{(2\pi )^3} \bar\psi_{K(1/2)}
\Bigl({\bf p}+\frac{2\epsilon_q }{E_{K(1/2)}+
M_{K(1/2)}}{\bf \Delta } \Bigr)\cr \cr
&&\times\sqrt{\frac{\epsilon_s(\Delta )+m_s}{2\epsilon_s(\Delta )}}
\Biggl\{\left[3(E_{K(1/2)}+M_{K(1/2)})
\frac{({\bf p}\cdot {\bf\Delta})}{p{\bf \Delta}^2}
+\frac{2p}{\epsilon_q(p)+m_q }\right]\cr \cr
&&\times\frac{\epsilon_s(\Delta )-m_s}
{2\epsilon_s(\Delta )[\epsilon_s(\Delta )+m_s]}
\biggl(M_B-\epsilon_b(p)-\epsilon_q(p)
+\frac{M_B-E_{K(1/2)}}{\epsilon_s(\Delta )+m_s}
\biggl[M_{K(1/2)}\cr \cr
&&
-\epsilon_s\Bigl({\bf p}+\frac{
2\epsilon_q }{E_{K(1/2)}+M_{K(1/2)}}{\bf \Delta }
\Bigr)- \epsilon_s\Bigl({\bf p}+
\frac{2\epsilon_q }{E_{K(1/2)}+M_{K(1/2)}}{\bf
\Delta } \Bigr) \biggr]\biggr)\cr\cr
&&-\frac{p}{2(\epsilon_q(p)+m_q) }\Biggl[
\frac{1}{\epsilon_b(\Delta )+m_b}
\Bigglb((M_B-\epsilon_b(p)-\epsilon_q(p))\cr\cr
&&\times \left(
3\frac{M_B+M_{K(1/2)}}{M_B-M_{K(1/2)}}-1\right)+\frac{M_B+M_{K(1/2)}} 
{M_B-M_{K(1/2)}} \biggl(M_{K(1/2)}\cr\cr
&&-
\epsilon_s\Bigl({\bf p}+\frac{
2\epsilon_q }{E_{K(1/2)}+M_{K(1/2)}}{\bf \Delta }
\Bigr) - \epsilon_s\Bigl({\bf p}+
\frac{2\epsilon_q }{E_{K(1/2)}+M_{K(1/2)}}{\bf
\Delta } \Bigr) \biggr)\Biggrb)\cr\cr
&&+\frac{3}
{2\epsilon_s(\Delta )}\Bigglb(\frac{M_B+M_{K(1/2)}}{M_B-M_{K(1/2)}}
(M_B-\epsilon_b(p)-\epsilon_q(p))+\Biggl(\frac{M_B+M_{K(1/2)}}{M_B-M_{K(1/2)}}\cr \cr
&&
-2\frac{\epsilon_s(\Delta )-m_s}{\epsilon_s(\Delta )+m_s}\Biggr)
\biggl(M_{K(1/2)}-
\epsilon_s\Bigl({\bf p}
+\frac{2\epsilon_q }{E_{K(1/2)}+M_{K(1/2)}}{\bf \Delta }
\Bigr)\cr\cr
&&- \epsilon_s\Bigl({\bf p}+
\frac{2\epsilon_q }{E_{K(1/2)}+M_{K(1/2)}}{\bf
\Delta } \Bigr)\biggr)\Biggrb) \Biggr]
\Biggr\} \psi_B({\bf p}),\\  \cr
\label{s2s}
\tilde s_{+}^{(2)S}(0) &=&\frac{1}{3} \sqrt{\frac{E_{K(1/2)}}{M_B}}
\frac{|{\bf \Delta}|}{E_{K(1/2)}+M_{K(1/2)}}
\int \frac{d^3p}{(2\pi )^3} \bar\psi_{K(1/2)}
\Bigl({\bf p}+
\frac{2\epsilon_q }{E_{K(1/2)}+M_{K(1/2)}}{\bf \Delta } \Bigr)\cr \cr
&&\times\sqrt{\frac{\epsilon_s(\Delta )+m_s}{2\epsilon_s(\Delta )}}
\Biggl\{\Biggl[-3(E_{K(1/2)}+M_{K(1/2)})
\frac{({\bf p}\cdot {\bf\Delta})}{p{\bf \Delta}^2}-
\frac{2p}{\epsilon_q(p)+m_q }\Biggr]\cr \cr
&&\times\frac{\epsilon_s(\Delta )-m_s}
{2\epsilon_s(\Delta )[\epsilon_s(\Delta )+m_s]}\Biggl[M_B-\epsilon_b(p)-\epsilon_q(p)
-\frac{M_B-E_{K(1/2)}}{\epsilon_s(\Delta )+m_s}
\bigglb(M_{K(1/2)}\cr\cr
&& -\epsilon_s\Bigl({\bf p}+\frac{
2\epsilon_q }{E_{K(1/2)}+M_{K(1/2)}}{\bf \Delta }
\Bigr)-  \epsilon_s\Bigl({\bf p}+
\frac{2\epsilon_q }{E_{K(1/2)}+M_{K(1/2)}}{\bf
\Delta } \Bigr) \biggrb)
\Biggr]\Biggr\} \psi_B({\bf p}),\\
\label{gpl}
g_{+}(0)&=&g_{+}^{(1)}(0)+
(1-\varepsilon)g_{+}^{(2)V}(0)+\varepsilon g_{+}^{(2)S}(0),\\
\label{g1}
g_{+}^{(1)}(0) &=&\frac{1}{\sqrt{3}} \sqrt{\frac{E_{K^*_2}}{M_B}}
\frac{M_{K_2^*}}{E_{K^*_2}+
M_{K_2^*}} \int \frac{d^3p}{(2\pi )^3} \bar\psi_{K_2^*}
\Bigl({\bf p}+
\frac{2\epsilon_q }{E_{K^*_2}+M_{K_2^*}}{\bf \Delta } \Bigr)\cr \cr
&&\times\sqrt{\frac{\epsilon_s(p+\Delta )+m_s}{2\epsilon_s(p+\Delta )}}
\sqrt{\frac{\epsilon_b(p )+m_b}{2\epsilon_b(p )}}
\Biggl\{-3(E_{K^*_2}+M_{K_2^*})
\frac{({\bf p}\cdot {\bf\Delta})}{p{\bf \Delta}^2}\cr \cr
&&\times\left[1+\frac{M_B-E_{K^*_2}}
{\epsilon_s(p+\Delta )+m_s } \right]
+\left[\frac{p}{\epsilon_q(p)+m_q }
-\frac{p}{\epsilon_s(p+\Delta )+m_s} \right] \cr \cr
&&\times\left[1+\frac{M_B-E_{K^*_2}}
{\epsilon_s(p+\Delta )+m_s }
-\frac{{\bf p}^2}{[\epsilon_s(p+\Delta )+m_s]
[\epsilon_b(p )+m_b]} \right]\Biggr\} \psi_B({\bf p}),\\  \cr
\label{g2v}
g_{+}^{(2)V}(0) &=&\frac{1}{\sqrt{3}} \sqrt{\frac{E_{K^*_2}}{M_B}}
\frac{M_{K_2^*}}{E_{K^*_2}+
M_{K_2^*}} \int \frac{d^3p}{(2\pi )^3} \bar\psi_{K_2^*}
\Bigl({\bf p}+\frac{2\epsilon_q }{E_{K^*_2}+
M_{K_2^*}}{\bf \Delta } \Bigr)\cr \cr
&&\times\sqrt{\frac{\epsilon_s(\Delta )+m_s}{2\epsilon_s(\Delta )}}
\Biggl\{3(E_{K^*_2}+M_{K_2^*})
\frac{({\bf p}\cdot {\bf\Delta})}{p{\bf \Delta}^2}
\frac{M_B-\epsilon_b(p)-\epsilon_q(p)}{2[\epsilon_s(\Delta)+m_s]}
-\frac{p}{\epsilon_q(p)+m_q }\cr \cr
&&\times\frac{1}{2[\epsilon_s(\Delta)+m_s]^2}\biggl((M_B+M_{K_2^*})
[M_B-\epsilon_b(p)-\epsilon_q(p)]
+(E_{K_2^*}+M_{K_2^*})\cr \cr
&&\times \biggl[M_{K_2^*}-
\epsilon_s\Bigl({\bf p}+\frac{
2\epsilon_q }{E_{K^*_2}+M_{K_2^*}}{\bf \Delta }
\Bigr)-  \epsilon_s\Bigl({\bf p}+
\frac{2\epsilon_q }{E_{K^*_2}+M_{K_2^*}}{\bf
\Delta } \Bigr) \biggr]\biggr)\Biggr\} \psi_B({\bf p}),\\  \cr
\label{g2s}
g_{+}^{(2)S}(0) &=&\frac{1}{\sqrt{3}} \sqrt{\frac{E_{K^*_2}}{M_B}}
\frac{M_{K_2^*}}{E_{K^*_2}+M_{K_2^*}}
\int \frac{d^3p}{(2\pi )^3} \bar\psi_{K_2^*}
\Bigl({\bf p}+
\frac{2\epsilon_q }{E_{K^*_2}+M_{K_2^*}}{\bf \Delta } \Bigr)\cr \cr
&&\times\sqrt{\frac{\epsilon_s(\Delta )+m_s}{2\epsilon_s(\Delta )}}
\Biggl\{\Biggl[-3(E_{K^*_2}+M_{K_2^*})
\frac{({\bf p}\cdot {\bf\Delta})}{p{\bf \Delta}^2}+
\frac{p}{\epsilon_q(p)+m_q }\Biggr]\cr \cr
&&\times\Biggl(\frac{\epsilon_s(\Delta )-m_s}
{2\epsilon_s(\Delta )[\epsilon_s(\Delta )+m_s]}
 \biggl[M_{K_2^*}-
\epsilon_s\Bigl({\bf p}+\frac{
2\epsilon_q }{E_{K^*_2}+M_{K_2^*}}{\bf \Delta }
\Bigr)\cr \cr
&&-  \epsilon_s\Bigl({\bf p}+
\frac{2\epsilon_q }{E_{K^*_2}+M_{K_2^*}}{\bf
\Delta } \Bigr) \biggr]
+\frac{M_B-E_{K^*_2}}{2[\epsilon_s(\Delta )+m_s]^2 }
 \biggl[M_B+M_{K_2^*}-
\epsilon_b(p)-\epsilon_q(p)\cr \cr
&&-\epsilon_s\Bigl({\bf p}+\frac{
2\epsilon_q }{E_{K^*_2}+M_{K_2^*}}{\bf \Delta }
\Bigr)-  \epsilon_s\Bigl({\bf p}+
\frac{2\epsilon_q }{E_{K^*_2}+M_{K_2^*}}{\bf
\Delta } \Bigr) \biggr]\Biggr)\Biggr\} \psi_B({\bf p}),
\end{eqnarray}
where the superscripts ``(1)" and ``(2)" correspond to contributions coming
from Figs.~1 and 2,
$S$ and $V$  mean the scalar and vector potentials in Eq.~(\ref{vconf}),
$\psi_{K^{**},B}$ are radial parts of the wave functions. 
Since $M_{K(1/2)}$ and $M_{K(3/2)}$ almost coincide with the physical 
axial-vector meson masses $M_{K^{(*)}_1}$ and $M_{K_1}$ we use
the latter for numerical calculations. The
recoil momentum and the energy of the ${K^{**}}$ meson are given by
\begin{equation}\label{delta}
\vert {\bf\Delta}\vert=\frac{M_B^2-M_{K^{**}}^2}{2M_B};
\qquad E_{K^{**}}=\frac{M_B^2+M_{K^{**}}^2}{2M_B}.
\end{equation}

\section{Results and discussion}
\label{sec:rd}

We can check the consistency of our resulting formulas by taking
the formal limit of   $b$ and $s$ quark masses going to
infinity.~\footnote{As it was noted above  such limit is
justified only for the $b$ quark.} In this limit according to
HQET \cite{vo2} the functions
\begin{equation}
  \label{eq:xi}
  \xi_F(w)=\frac{2\sqrt{M_BM_{K_2^*}}}{M_B+M_{K_2^*}}g_{+}(w)=
\frac{2\sqrt{M_BM_{K_1}}}{M_B-M_{K_1}}\frac{\sqrt{6}}{w+1}\tilde t_{+}(w)
\end{equation}
and
\begin{equation}
\label{eq:xie}
 \xi_E(w)=
\frac{2\sqrt{M_BM_{K_1^*}}}{M_B-M_{K_1^*}}\tilde s_{+}(w), \qquad
w=\frac{M^2_B+M^2_{K^{**}}-k^2}{2M_B M_{K^{**}}},
\end{equation}
should coincide
with the Isgur-Wise functions $\tau(w)$ and $\zeta(w)$ for semileptonic $B$
decays to orbitally excited $D$ mesons, $B\to
D^{**}e\nu$. Such semileptonic decays have been considered by us
in Ref.~\cite{orb}. Taking the formal limit $m_b\to\infty$, $m_s\to\infty$ in
Eqs.~(\ref{tpl})--(\ref{g2s}) and using definitions (\ref{eq:xi}),
(\ref{eq:xie}) we find
\begin{eqnarray}
\label{xiF}
\xi_F(w)&=&\sqrt{\frac23}\frac{1}{(w+1)^{3/2}}
\int \frac{d^3p}{(2\pi )^3} \bar\psi_{K(3/2)}
\Bigl({\bf p}+\frac{2\epsilon_q }{M_{K(3/2)}(w+1)}{\bf \Delta }\Bigr)\cr\cr
&& \times\left[-3M_{K(3/2)}(w+1)
\frac{({\bf p}\cdot {\bf\Delta})}{p{\bf \Delta}^2}
+\frac{p}{\epsilon_q(p)+m_q}\right] \psi_B({\bf p}),\\ \cr
\label{xiE}
\xi_E(w)&=&\frac{\sqrt{2}}{3}\frac{1}{(w+1)^{1/2}}
\int \frac{d^3p}{(2\pi )^3} \bar\psi_{K(1/2)}
\Bigl({\bf p}+\frac{2\epsilon_q }{M_{K(1/2)}(w+1)}{\bf \Delta }\Bigr)\cr\cr
&& \times\left[-3M_{K(1/2)}(w+1)
\frac{({\bf p}\cdot {\bf\Delta})}{p{\bf \Delta}^2}
-2\frac{p}{\epsilon_q(p)+m_q}\right] \psi_B({\bf p}).
\end{eqnarray}
It is easy to verify that the equalities
$\xi_F= \tau$ and $\xi_E=\zeta$ are implemented in our
model if we also use the
expansion in $(w-1)/(w+1)$ ($w$ is a scalar product of
four-velocities of the initial and final mesons), which is small
for the $B\to D^{**}e\nu$ decay \cite{orb}. It is important to note
that last terms in the square brackets of the expressions for the functions
$\xi_F(w)$ (\ref{xiF}) and $\xi_E(w)$ (\ref{xiE}) result from the wave
function transformation (\ref{wig}) associated with the relativistic rotation of the
light quark spin (Wigner rotation) in
passing to the moving reference frame. These terms are numerically important
and lead to the suppression of the $\xi_E$ form factor compared to
$\xi_F$. Note that if we
applied a simplified non-relativistic quark model \cite{vo,vo2}
these important contributions would be missing. Neglecting further the
small difference between the wave functions $\psi_{K(1/2)}$ and
$\psi_{K(3/2)}$, the following relation between $\xi_F$ and
$\xi_E$ would be obtained \cite{vo2}
\begin{equation}\label{taunr}
\xi_E(w)=\frac{w+1}{\sqrt{3}}\xi_F(w).
\end{equation}
However, we see that this relation is violated if the relativistic
transformation properties of the wave function are taken into
account.

The relations between the form factors of heavy-to-light semileptonic
and rare radiative $B$ decays emerging in the large recoil limit
\cite{cbf} are satisfied in our model \cite{gf,fgm}.

Using Eq.~(\ref{gpl})
to calculate the ratio of  the form factor $g_+(0) $ in the
infinitely heavy $b$ and $s$ quark limit to the same form factor
in the leading order of  expansions in inverse powers of the
heavy $b$ quark mass and large recoil momentum $|{\bf\Delta}|$
we find that it is equal to $M_B/\sqrt{M_B^2+M_{K^*_2}^2}\approx
0.965$. The corresponding ratio of form factors of the exclusive
rare radiative $B$ decay to the vector $K^*$ meson $F_1(0)$ (see
Eq.~(23) of Ref.~\cite{gf}) is equal to
$M_B/\sqrt{M_B^2+M_{K^*}^2}\approx 0.986$. Therefore we conclude
that the form factor ratios  $g_+(0)/F_1(0)$ in the leading
order of these expansions differ by the factor
$\sqrt{M_B^2+M_{K^*}^2}/\sqrt{M_B^2+M_{K^*_2}^2}\approx 0.98$.
This is the consequence of the relativistic dynamics leading to
the effective expansion in inverse powers of the $s$ quark energy
$\epsilon_s(p+\Delta)=\sqrt{({\bf p+\Delta})^2+m_s^2}$, which is
high in one case due to the large $s$ quark mass and in the
other one due to the large recoil momentum ${\bf \Delta}$.  As a
result both expansions give similar final expressions in the
leading order. Thus we can expect that the ratio $r$ of the $B$ branching
fractions to the tensor $K^*_2$ and vector $K^*$ mesons
in our calculations should be close to the one
found in the infinitely heavy $s$ quark limit \cite{vo}.

The results of numerical calculations using formulas
(\ref{drate1})--(\ref{rk2}), (\ref{eq:fff}), (\ref{tpl})--(\ref{g2s}) for
$\varepsilon=-1$ are given in Table~\ref{tb}. There we also show
our previous predictions for the $B\to K^*\gamma$ decay \cite{gf}.
Our results are confronted  with other theoretical calculations
\cite{a,aom,as,vo} and recent experimental data \cite{cleo2}.
The QCD sum rules predict (with 20\% uncertainty) \cite{b} $BR(B\to K^*\gamma)
= 4.4\times 10^{-5}\times(1+8\%)$, where the second term in the
brackets is the estimate of the $1/m_c^2$ terms contribution. We
find a good agreement of our predictions for decay rates with the
experiment and estimates of Ref.~\cite{vo} for the measured decay
rates $B\to K^*\gamma$ and $B\to K_2^*\gamma$. Other theoretical
calculations substantially disagree with data either for $B\to
K^*\gamma$ \cite{a,as} or for $B\to K^*_2\gamma$ \cite{aom} decay
rates. Let us note that one of the main reasons of the 
too small values
for $B\to K^*\gamma$ decay
rates in quark models \cite{a,as} is the use of
the nonrelativistic expression for the momentum of the
final meson in the
argument of the wave function overlap \cite{gf}.
As a result our predictions and those of Ref.~\cite{vo}
for the ratio $r$ are well consistent with
experiment, while the $r$ estimates of \cite{a,as} and \cite{aom}
are several times larger than the experimental value (see
Table~\ref{tb}). As it was argued above, it is not accidental
that $r$ values in our and Ref.~\cite{vo} approaches are close.
The agreement of both predictions for branching fractions could be
explained by some specific cancellation of finite $s$ quark mass
effects and relativistic corrections which were neglected in
Ref.~\cite{vo}. Though our numerical results for the measured decay
rates agree with
Ref.~\cite{vo}, we believe that our analysis is more consistent
and reliable. We do not use the ill-defined limit $m_s\to\infty$,
and our quark model consistently takes into account main
relativistic effects, for example, the Lorentz
transformation of the wave function of the final $K^{**}$ meson
(see Eq.~(\ref{wig})). Such a transformation turns out to be very
important and leads to the substantial reduction of
$B\to K^*_1(1270)\gamma$ decay rate in our model.
We see from Table~\ref{tb} that
our model predicts for the ratio $BR(B\to K_1^*(1270)\gamma)/BR(B\to
K_1(1430)\gamma)$ the value $0.7\pm 0.3$ while Ref.~\cite{vo} gives for
this ratio a
considerably larger value $\sim 2$, which is the consequence
of the nonrelativistic quark model relation (\ref{taunr}) between form factors
$\xi_F$ and $\xi_E$. Thus experimental
measurement of $BR(B\to K_1^*(1270)\gamma)$ can discriminate between
these predictions.

\section{Conclusions}
\label{sec:conc}

In this paper we have investigated rare radiative $B$ decays to
orbitally excited $K^{**}$ mesons in the framework of the relativistic
quark model. The large value of the recoil momentum $|{\bf
\Delta}|\sim m_b/2$ makes relativistic effects to play a
significant role and strongly
increases the energy of the final meson. This effect considerably
simplifies the analysis since it allows to make an expansion both in
inverse powers of the large $b$ quark mass and in the large recoil
momentum of the light final meson. Such an expansion has more firm
theoretical grounds than the previously used expansion
in inverse powers of the $s$ quark mass \cite{vo,aom},
which is not heavy enough.
We carried out this expansion up to the second order and calculated
resulting form factors in our relativistic quark model. Rare radiative
$B$ decays to axial-vector $K_1^{(*)}$ and tensor $K_2^*$ mesons have
been considered. It was found that relativistic effects
substantially influence decay form factors. Thus, the Wigner
rotation of the light quark spin gives an important contribution,
which leads to the suppression of the $B\to K^*_1(1270)\gamma$ decay
rate. In the nonrelativistic quark model, where these effects are missing,
the ratio of branching fractions $BR(B\to K^*_1(1270)\gamma)/BR(B\to
K_1(1400)\gamma)$  is equal to $2$, while in our model it is
substantially smaller and equal to $0.7\pm 0.3$. It will be very
interesting to test this conclusion experimentally.

Our predictions for the branching fractions $B\to K^*\gamma$ and $B\to
K^*_2\gamma$ as well as their ratio are in a good agreement with recent
CLEO data \cite{cleo2}.

\acknowledgements
The authors express their gratitude to P. Ball, A. Golutvin,
M. Mueller-Preussker, and V. Savrin for discussions and D. Jaffe for
initiating this study. D.E. acknowledges the support provided
to him by the Ministry of Education and Science and Technology 
of Japan  (Monkasho) for his
work at RCNP of Osaka University. Two of us (R.N.F and V.O.G.)
were supported in part by the {\it Deutsche
Forschungsgemeinschaft} under contract Eb 139/2-1, {\it
 Russian Foundation for Fundamental Research} under Grant No.\
00-02-17768 and {\it Russian Ministry of Education} under Grant
No. E00-3.3-45.

\begin{table}
\caption{Theoretical predictions and experimental data for the
branching fractions ($\times 10^{-5}$)  and their
ratios $R_{K^*}\equiv {BR(B\to K^*\gamma)}/{BR(B\to
X_s\gamma)}$,  $R_{K_i^{(*)}}\equiv{BR(B\to
K_i^{(*)}\gamma)}/{BR(B\to X_s\gamma)}$ ($i=1,2$), $r\equiv{BR(B\to
K_2^*\gamma )}/{BR(B\to K^*\gamma)}$. Our values for the $B\to
K^*\gamma$ decay are taken from Ref.~\protect\cite{gf}.}
\label{tb}
\begin{tabular}{ccccccc}
Value&our&Ref.~\cite{a}&Ref.~\cite{aom}&Ref.~\cite{as}
&Ref.~\cite{vo}&Exp.  \cite{cleo2}\\
\hline
$BR(B\to K^*(892)\gamma)$& $4.5\pm1.5$
& 1.35 & $1.4 - 4.9$&$0.5-0.8$ &$4.71\pm1.79$
&$4.55^{+0.72}_{-0.68}\pm0.34^a $ \\
 & & & & & & $3.76^{+0.89}_{-0.83}\pm0.28^b$  \\
$R_{K^*}$ (\%) &$15\pm3$&4.5 &$3.5 - 12.2$&$1.6-2.5$ &$16.8\pm6.4$ &\\
\hline
$B\to K_0^*(1430)\gamma$ & & & forbidden & & & \\
\hline
$BR(B\to K_1^*(1270)\gamma)$&$0.45\pm 0.15$& 1.1&
$1.8-4.0$&$0.3-1.4$ &$1.20\pm 0.44$& \\
$R_{K_1^*}$ (\%) & $1.5\pm 0.5$ & 3.8& $4.5-10.1$&$0.9-4.5$ &$4.3\pm 1.6$&\\
\hline
$BR(B\to K_1(1400)\gamma)$&$0.78\pm 0.18$& 0.7&
$2.4-5.2$&$0.1-0.6$ &$0.58\pm 0.26$& \\
$R_{K_1}$ (\%) & $2.6\pm 0.6$ & 2.2& $6.0-13.0$&$0.4-2.0$ &$2.1\pm 0.9$&\\
\hline
$BR(B\to K_2^*(1430)\gamma)$&$1.7\pm0.6$
& 1.8 & $6.9 - 14.8$&$0.4-1.0$
& $1.73\pm0.80$& $1.66^{+0.59}_{-0.53}\pm0.13$ \\
$R_{K_2^*}$ (\%)
& $5.7\pm1.2$ & 6.0 & $17.3 -37.1$ &$1.3-3.2$ &$6.2\pm2.9$ & \\
$r$
& $0.38 \pm 0.08$ & 1.3 & $3.0 - 4.9$&$0.8-1.3$ & $0.37\pm 0.10$
& $0.39^{+0.15}_{-0.13}$\\
\end{tabular}

{\small $^a$ $B^0\to K^{*0}\gamma$

$^b$ $B^+\to K^{*+}\gamma$}
\end{table}

\begin{figure}
\unitlength=0.9mm
\begin{picture}(150,150)
\put(10,100){\line(1,0){50}}
\put(10,120){\line(1,0){50}}
\put(35,120){\circle*{5}}
\multiput(32.5,130)(0,-10){2}{\begin{picture}(5,10)
\put(2.5,10){\oval(5,5)[r]}
\put(2.5,5){\oval(5,5)[l]}\end{picture}}
\put(5,120){\large$b$}
\put(5,100){\large$\bar q$}
\put(5,110){\large$B$}
\put(65,120){\large$s$}
\put(65,100){\large$\bar q$}
\put(65,110){\large$K^{**}$}
\put(43,140){\large$\gamma$}
\put(0,85){\small FIG. 1. Lowest order vertex function $\Gamma^{(1)}$
corresponding to Eq.~(\ref{gam1}). }
\put(10,20){\line(1,0){50}}
\put(10,40){\line(1,0){50}}
\put(25,40){\circle*{5}}
\put(25,40){\thicklines \line(1,0){20}}
\multiput(25,40.5)(0,-0.1){10}{\thicklines \line(1,0){20}}
\put(25,39.5){\thicklines \line(1,0){20}}
\put(45,40){\circle*{1}}
\put(45,20){\circle*{1}}
\multiput(45,40)(0,-4){5}{\line(0,-1){2}}
\multiput(22.5,50)(0,-10){2}{\begin{picture}(5,10)
\put(2.5,10){\oval(5,5)[r]}
\put(2.5,5){\oval(5,5)[l]}\end{picture}}
\put(5,40){\large$b$}
\put(5,20){\large$\bar q$}
\put(5,30){\large$B$}
\put(65,40){\large$s$}
\put(65,20){\large$\bar q$}
\put(65,30){\large$K^{**}$}
\put(33,60){\large$\gamma$}
\put(90,20){\line(1,0){50}}
\put(90,40){\line(1,0){50}}
\put(125,40){\circle*{5}}
\put(105,40){\thicklines \line(1,0){20}}
\multiput(105,40.5)(0,-0.1){10}{\thicklines \line(1,0){20}}
\put(105,39,5){\thicklines \line(1,0){20}}
\put(105,40){\circle*{1}}
\put(105,20){\circle*{1}}
\multiput(105,40)(0,-4){5}{\line(0,-1){2}}
\multiput(122.5,50)(0,-10){2}{\begin{picture}(5,10)
\put(2.5,10){\oval(5,5)[r]}
\put(2.5,5){\oval(5,5)[l]}\end{picture}}
\put(85,40){\large$b$}
\put(85,20){\large$\bar q$}
\put(85,30){\large$B$}
\put(145,40){\large$s$}
\put(145,20){\large$\bar q$}
\put(145,30){\large$K^{**}$}
\put(133,60){\large$\gamma$}
\put(0,5){\makebox[14cm][s]{\small FIG. 2. Vertex function $\Gamma^{(2)}$
corresponding to Eq.~(\ref{gam2}). Dashed lines represent the}}
\put(0,0) {\makebox[14cm][s]{\small  interaction operator ${\cal V}$ in
Eq.~(\ref{qpot}). Bold lines denote the negative-energy part of the} }
\put(0,-5){\small quark propagator. }

\end{picture}
\end{figure}

\end{document}